\documentclass[twocolumn,10pt]{revtex4}%
\usepackage[paperwidth=210mm,paperheight=297mm,centering,hmargin=2cm,vmargin=2.5cm]{geometry}
\usepackage{amsfonts}
\usepackage{amsmath}
\usepackage{amssymb}
\usepackage{bm}
\usepackage{graphicx}
\usepackage{color}
\usepackage{soul}
\usepackage{epsfig}%
\usepackage[dvipsnames]{xcolor} 
  \definecolor{bleu_cite}{RGB}{0,0,255}
\usepackage[colorlinks=true,allcolors = black,citecolor=bleu_cite,  ]{hyperref} 
\usepackage{natbib}

\begin{document}
\title{Observation of Algebraic Time Order for Two-Dimensional Dipolar Excitons}

\author{Suzanne Dang$^1$, Marta Zamorano$^{1}$, Stephan Suffit$^{2}$, Kenneth West$^3$, Kirk  Baldwin$^3$, Loren Pfeiffer$^3$, Markus Holzmann$^4$ and Fran\c{c}ois Dubin$^{1}$} 
\affiliation{$^1$ Institut des Nanosciences de Paris, CNRS and Sorbonne Universit{\'e}, 4 pl. Jussieu,
75005 Paris, France}
\affiliation{$^1$ Laboratoire de Materiaux et Phenomenes Quantiques, Universite Paris Diderot, Paris}
\affiliation{$^3$ PRISM, Princeton Institute for the Science and Technology of Materials, Princeton Unviversity, Princeton, NJ 08540}
\affiliation{$^4$ Univ. Grenoble Alpes, CNRS, LPMMC, 3800 Grenoble, France}

\begin{abstract}

\end{abstract}

\maketitle
\textbf{Emergence of algebraic quasi-long-range order is a key feature of superfluid
phase transitions at two dimensions \cite{Berenzinskii,KT}. For this reduced dimensionality interactions prevent Bose-Einstein condensation with true long range order, at any finite temperature \cite{Mermin-Wagner}. Here, we report the occurence of algebraic order in a strongly interacting quantum liquid formed by dipolar excitons confined in a bilayer semiconductor heterostructure \cite{Combescot_ROPP}. We observe a transition from exponential to algebraic decay of the excitons temporal coherence, accompanied by a universal scaling behaviour of the equation of state. Our results provide strong evidence for a Berezinskii-Kosterlitz-Thouless (BKT) transition in a multi-component boson-like system governed by strong dipolar interactions.}

Dipolar quantum gases constitute a versatile playground to explore exotic collective phenomena emerging in a strong interaction regime \cite{Maciej_2011}. In two dimensions, important correlation effects are predicted for moderate and high densities  \cite{Filinov10}, 
causing substantial deviations from dilute gas behavior.
Very recently, supersolid properties \cite{Tanzi_19,Bottcher_19,Chomaz_19} have been reported for ultra-cold atoms with a large permanent magnetic dipole. In semiconductors, even stronger dipolar interactions are accessible, using  bilayer heterostructures where electrons and holes are spatially separated but bound by Coulomb attraction \cite{Combescot_book}. Excitons are thus formed, characterised by a giant and well oriented electric dipole moment \cite{Combescot_ROPP}. Recent experiments have reported that such dipolar excitons can realize a two dimensional quantum gas in GaAs bilayers \cite{High_2012,Anankine_2017,Nandi_2012,Dang_2019,Rapaport_2019}. 

Here, we first quantify thermodynamically the quasi-condensate crossover region of two-dimensional dipolar excitons. For that, we experimentally determine the equation of state
and density fluctuations. Whereas apparent scale invariance for these observables has been measured with ultracold atoms, even away from criticality \cite{Yefsah_2011,SaintJalm19,Hung_2011}, here we report a net violation confirming the strongly interacting character of our dipolar fluid. Also, we show that a universal behaviour \cite{Prokofev2002,Hung_2011} is restored within a large fluctuating region around the quasi-condensate cross-over. Importantly, close to the transition we observe a qualitative change in the excitons temporal phase coherence, from an exponential to an algebraic decay, revealing the buildup of a quasi-long range ordered phase.
The exponent of the algebraic decay, $\eta$, decreases with increasing phase space density, and is
compatible with $\eta_c\simeq 0.25$ predicted by the BKT theory \cite{Berenzinskii,KT,Nelson} at criticality.

Our experiments start by injecting optically dipolar excitons in a 20 $\mu$m wide trap of an ultra-pure GaAs double quantum well. This excitation lasts 100 ns, repeated at a frequency of 1.5 MHz, while  at a variable delay after termination of the optical loading we analyse the photoluminescence reemitted by the trapped gas and which reflects directly fundamental properties of the excitonic cloud \cite{Combescot_ROPP,Combescot_book}. Indeed, the energy of the photoluminescence E$_X$ provides the density in the trap $n$(\textbf{r}) since E$_X$ scales as (E$_\mathrm{t}$(\textbf{r})+$u_0$$n$(\textbf{r})),  the second term reflects the strength of repulsive dipolar interactions between excitons \cite{Ivanov_2010,Schindler_2008,Rapaport_2009}, and the first term the profile of the trapping potential. Additional information is gained from the photoluminescence time coherence, which characterizes the collisional dynamics and thus quantifies the excitons temporal phase coherence.

In our system, excitons are characterised by a dipole moment $d$, oriented perpendicular to the
GaAs bilayer, with a magnitude controlled by the spatial separation between the two quantum wells.
The effective exciton-exciton interaction potential is well described \cite{Ivanov_2010,Schindler_2008,Rapaport_2009} by  
V$_\mathrm{eff}$(\textbf{r})=$(d_\mathrm{eff})^2$/$|\textbf{r}|^3$, where $d_\mathrm{eff}=d\sqrt{f}$ contains a screening amplitude $f$ that weakly depends on the exciton density in the dilute regime ($f$ being of the order of 0.2 for $n\sim$10$^{10}$ cm$^{-2}$ at low temperatures) and may be adapted to also include
correlation effects at intermediate and high densities \cite{Filinov10}.
In two dimensions, the dipolar potential  V$_\mathrm{eff}$  is still sufficiently short ranged, so that
asymptotic scattering properties at large distances are characterized by 
a dimensionless number $\tilde{g}$ \cite{Popov}.

\begin{figure}\label{fig1}
\centerline{\includegraphics[width=.5\textwidth]{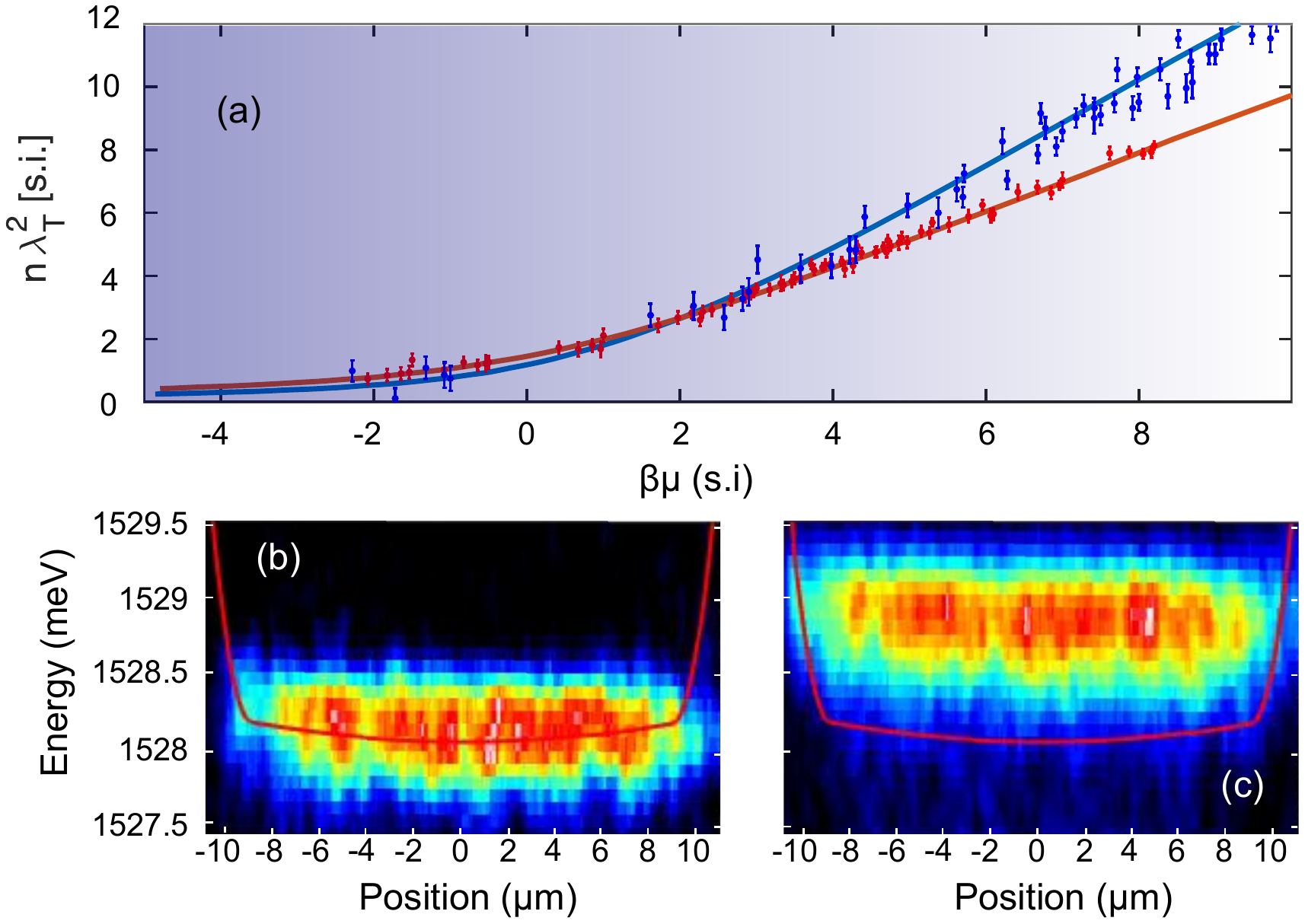}}
\caption{\textbf{Equation of state of the trapped gas} (a) Phase space density $D$=$n\lambda_T^2$ measured at $T$=340 mK (blue) and 750 mK (red), together with the results of Monte-Carlo calculations for $\tilde{g}$=4 and 6 respectively (solid lines). $D$ is obtained by analysing the profile of the photoluminescence energy across the trap from a statistical ensemble of 20 repetitions for every experimental conditions. Experimentally, the density in the trap is varied by scanning the delay to the loading phase. It is about 2$\cdot$10$^{10}$ cm$^{-2}$ at the center when the delay is set to 150 ns (c) and around 10$^{9}$ cm$^{-2}$ when it is set to 370 ns (b). In the latter case the photoluminescence energy reproduces the profile of the trapping potential depicted by the solid red line. Error bars display the statistical deviations of our measurements.}
\end{figure}

In Figure 1.a, we show the phase space densities $D(\beta\mu)$=$n \lambda^2_\mathrm{T}$ 
as a function of $\beta\mu$=$\mu/k_\mathrm{B}T$,
$\lambda_\mathrm{T}$=h/$\sqrt{2\pi mk_\mathrm{B}T}$ being the de Broglie thermal wavelength, $m$ the
exciton mass, $T$ the temperature, and $\mu$ the chemical potential.
 These measurements are
obtained from the density profiles of our excitonic cloud based on the local density approximation \cite{Dang_2019}. Unlike quasi-two-dimensional atomic gases  probed in the weak interaction regime  \cite{Yefsah_2011,SaintJalm19,Hung_2011}, a net violation of scale invariance in our system is directly evidenced by comparing experiments at $T$=340 and 750 mK. Indeed, $D$ does not obey a unique function  of
 $\beta\mu$. Instead, the distinct slopes observed between these two temperatures when $\beta\mu\gtrsim$0 show that $\tilde{g}$ decreases with temperature,
since $D(\beta\mu)\sim\beta\mu/\tilde{g}$ is expected in the Thomas-Fermi limit of a quasi-condensate. 
Comparison with classical field Monte Carlo calculations taking into account the exciton's composite nature
with four internal spin degrees of freedom \cite{Dang_2019,HCK} 
indicates that $\tilde{g}$ increases from about 4 to 6 between 340 and 750 mK.

To further quantify this temperature dependence, we studied the exciton density variance $\sigma^2$(\textbf{r}) that provides an independent and direct measure of $\tilde{g}$ in the quasi-condensed regime \cite{Hung_2011}. Figure 2.a shows the scaled variance $\sigma^2$$\lambda^2_\mathrm{T}$ per $\mu m^{2}$ at variable temperatures, within the hydrodynamic regime where $D$ scales linearly with $\beta\mu$. We note that  $\sigma^2$$\lambda^2_\mathrm{T}$  weakly varies with $\beta\mu$ and approaches a mean value which mostly depends on $T$. Let us then underline that our experiments rely on cumulative averaging for each experimental settings. The measured local variance  $\sigma^2$(\textbf{r}) then corresponds to  the mean square deviation of  the total number of excitons at a considered position $\textbf{r}$. Furthermore, our optical resolution ($\sim 1 \mu$m) largely exceeds the mean-field coherence length $(2 \tilde{g} n)^{-1/2}$ ($\sim 25$nm). We then determine the fluctuations of the local  number of excitons so that $\sigma^2 \lambda^2_\mathrm{T}$ scales as 1/$\tilde{g}$ in the quasi condensed regime where  $D$ scales linearly with $\beta\mu$ (see Methods).

\begin{figure}\label{fig2}
\centerline{\includegraphics[width=.5\textwidth]{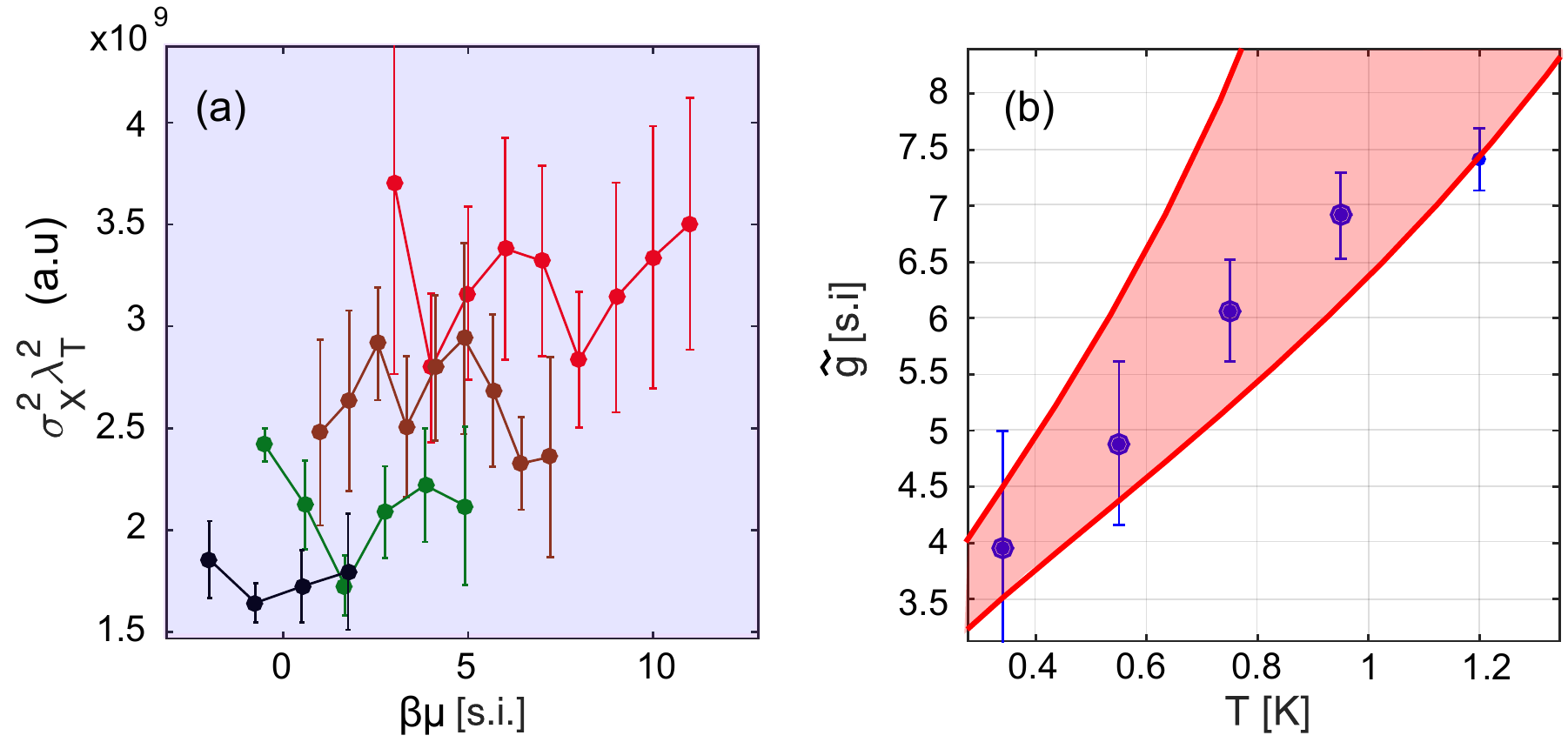}}
\caption{\textbf{Density fluctuations in the trap} (a) Rescaled density fluctuations $\sigma^2\lambda_T^2$ per $\mu m^2$ measured at 340 mK (red), 550 mK (brown), 950 mK (green), 1.2 K (black). Data points are obtained by evaluating the variance for a set of 20 repetitions for every experimental settings, $\sigma^2\lambda_T^2$  is then computed in intervals $\beta\mu\sim$1 to reach higher precision. (b) Temperature scaling of $\tilde{g}$ extracted by averaging the data shown in (a) for each bath temperature (blue). The red shaded area marks the logarithmic scaling of $\tilde{g}$ with temperature
expected for thermal two-body collisions with the effective dipolar exciton-exciton interaction \cite{Dang_2019}.}
\end{figure}

Averaging  for each temperature $T$ the rescaled variance shown in Fig.2.a, we deduce the dependence of $\tilde{g}$ as a function of $T$, imposing $\tilde{g}$=4 at 340 mK according to the analysis of the equation of state. Figure 2.b displays the thus obtained values, which are consistent with those from the Thomas-Fermi behaviour used for the equation of state (Fig.1), as expected from thermodynamics. Importantly, the temperature scaling of $\tilde{g}$ is consistent with a logarithmic dependence characteristics of two dimensional scattering processes.

\begin{figure}\label{fig3}
\centerline{\includegraphics[width=.5\textwidth]{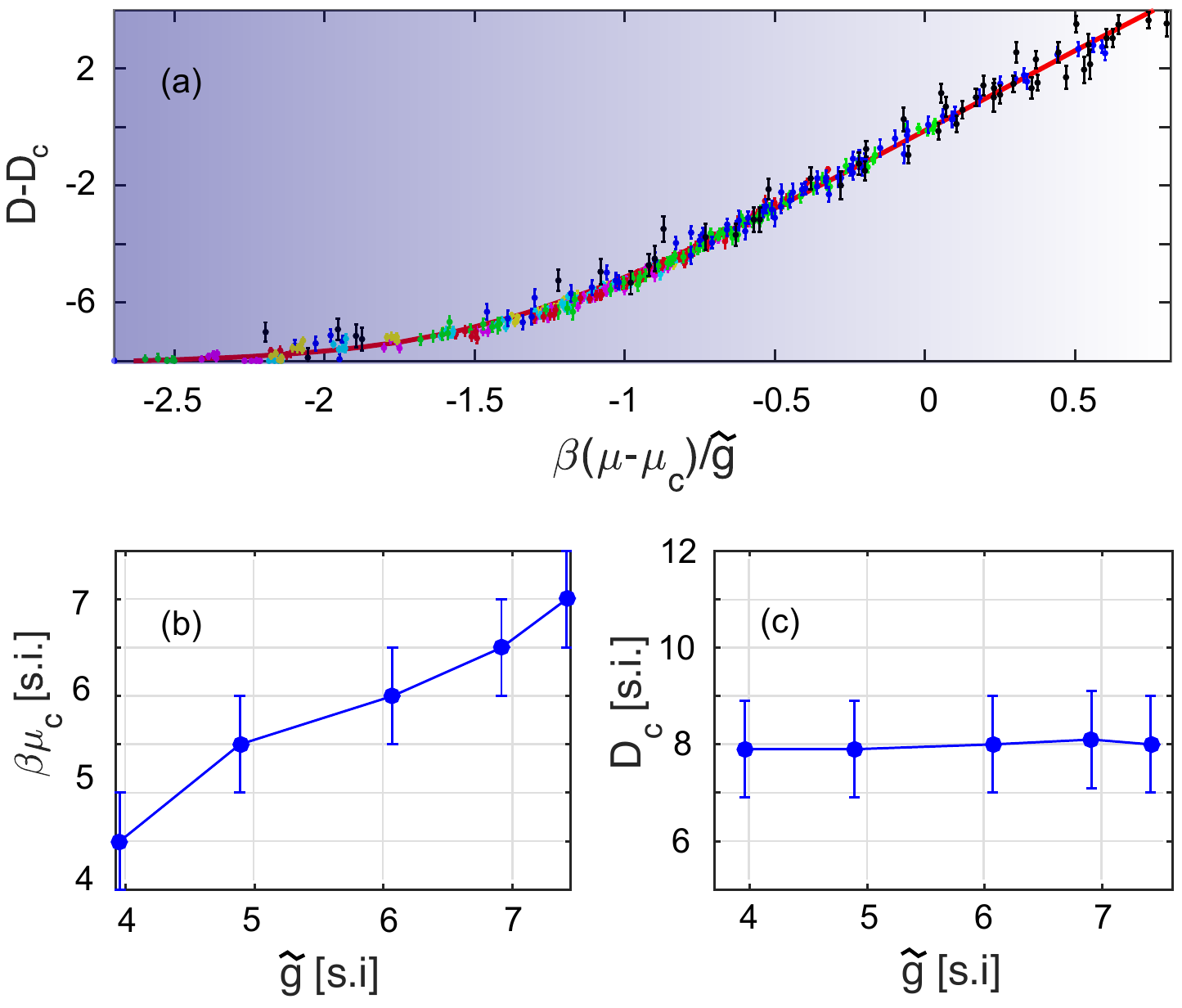}}
\caption{\textbf{Universality of the rescaled equations of state} (a) Equations of state ($D-D_c$) in rescaled units $\beta(\mu-\mu_c)/\tilde{g}$, for experiments realised at 340, 550, 750, 950 mK, 1.2  and 1.5 K (black, blue, green, red, cyan, magenta respectively). The solid red line displays the scaling predicted by Monte-Carlo calculations. 
The resulting values of the critical chemical potential, $\beta\mu_c$, and 
of the critical phase space density, $D_c$, are shown in (b) and (c), respectively.}
\end{figure}

Despite a broken scale invariance, scaling behaviours may be retrieved by the universal
variations expected close to a continuous phase transition such as the BKT transition. 
In particular, from the $\phi^4$  theory in two dimensions \cite{Prokofev2002,Pnas}, we expect 
the phase space density to show a universal scaling, 
($D-D_c$)=$F(\beta(\mu-\mu_c)/\tilde{g}$) inside the fluctuating region, $\beta|\mu-\mu_c|<\tilde{g}$. Here, $F$ is a generic function while $D_c$ and $\mu_c$ are the critical density and chemical potential which remove non-universal contributions with regular (smooth) temperature and density dependence.
Having determined the variation of $\tilde{g}$ with temperature, we can probe the universality
of the equation of state close to the quasi-condensation
crossover. To this aim, we transformed the equation of state by converting $\beta\mu$ into $\beta\mu/\tilde{g}$
 and then searched the values of $D_c$ and $\mu_c$ such that a unique variation emerges for all temperatures studied experimentally.

 As shown in Figure 3, all rescaled equations of state then collapse on a single
scaling curve, in striking contrast with the unscaled data displayed in Fig.1.a. 
The corresponding values for the critical chemical potential $\mu_c$ show a linear dependence on $\tilde{g}$,
whereas $D_c \approx 8$ remains roughly constant. The latter value also corresponds to the
onset of quasi long-range order in the spatial coherence \cite{Dang_2019}, as well as the temporal coherence discussed below. In contrast to the BKT transition in weakly interacting atomic gases \cite{Yefsah_2011}, continuously approaching the BEC transition for vanishing $\tilde{g}$ in a harmonic trap \cite{Fletcher15}, for our dipolar gas the transition takes place well above onset of the Thomas-Fermi region of the density profile where
$D \propto \beta \mu$. 
Also, the saturation of $D_c$ confirms a strongly correlated regime away from the dilute gas limit \cite{Filinov10}. In the crossover region, i.e. for $D\sim D_c$, we actually deduce that the
first sound velocity $v_s$=$\sqrt{\hbar^2\tilde{g}n}/m$ is about 10$^4$ m.s$^{-1}$, i.e. about three orders of magnitude smaller than for exciton-polaritons \cite{Amo_2009}. Therefore, dipolar exciton fluids provide an ideal probe of BKT physics in the hydrodynamic regime dominated by collisions between low-energy quasi-particle excitations.

\begin{figure}[h!]\label{fig4}
\centerline{\includegraphics[width=.5\textwidth]{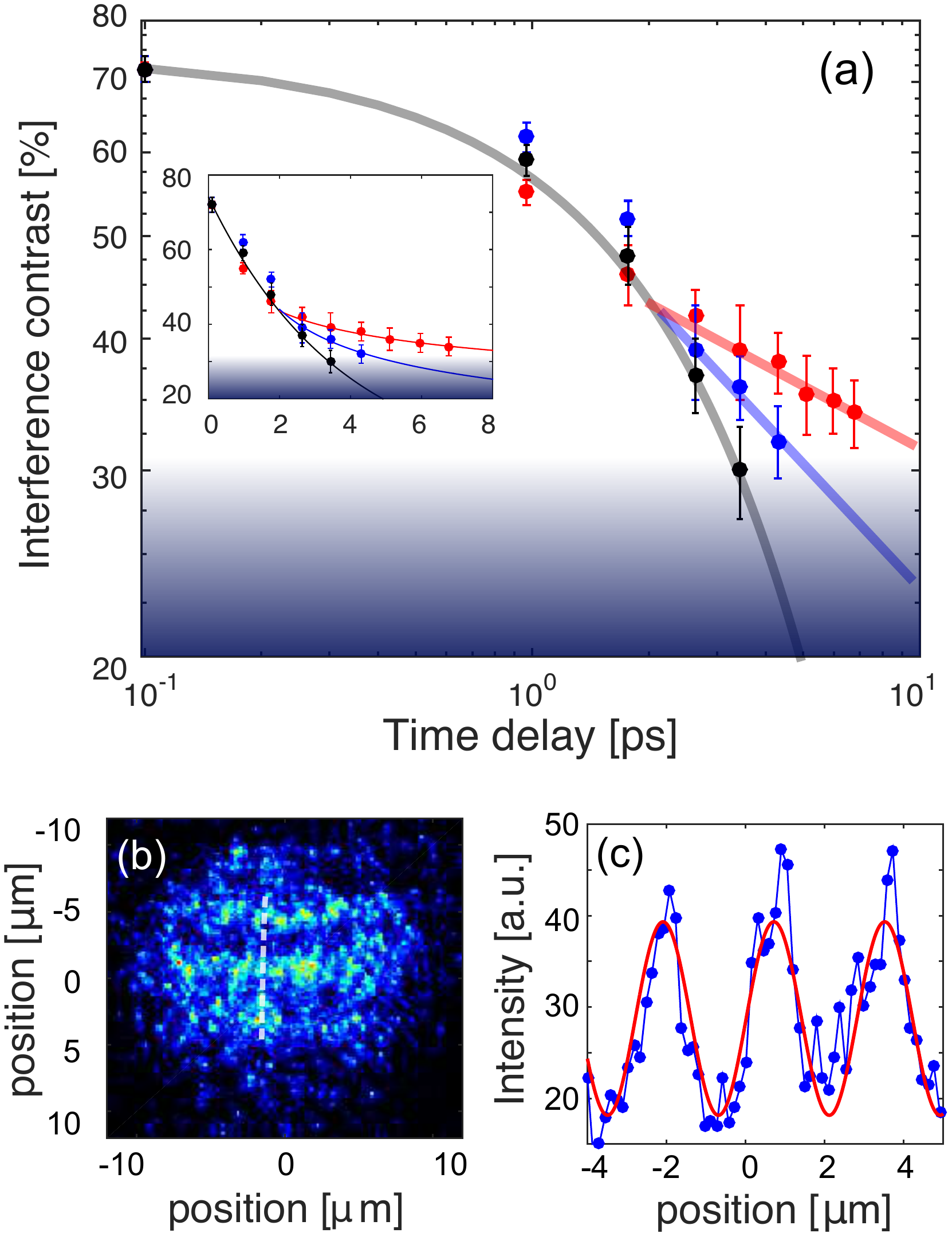}}
\caption{\textbf{Temporal coherence versus phase-space density} (a) Amplitude of the first-order temporal coherence $|g^{(1)}|$ as a function of the time delay for a phase-space
density below threshold ($D= 3$ in black), at threshold ($D\sim$7 in blue),
and above threshold ($D=11$ in red), in log-log scale. The inset displays the same data in linear scale. The black curve shows an exponential decay with a time constant equal to 3.8 ps while the blue and red lines show
an algebraic decay $t^{ˆ'-\eta}$ with $\eta=0.4$ and 0.2, blue and red respectively, for comparison. The blue shaded area marks the minimum contrast possibly detected for the signal to noise ratio of our experiments. (b) Interference pattern measured for $D=11$ and for a time delay set to $5$ ps. The panel (c) displays the interference signal across the center of the trap (dashed line in (b)), together with a sinusoidal fit to the data leading to a contrast of 35 \%.}
\end{figure}

We studied the temporal coherence of trapped excitons to independently assess the parameter region of the quasi-condensate crossover. Around $D_c$ we expect that temporal coherence is maintained at long times, thus reflecting directly the building up of spatial order in the system  \cite{Prokofev2018,Caputo_2018}. For excitonic systems, temporal coherence is easily accessed, since it is mapped out by the emitted photoluminescence, which quantifies the coherence time $\tau_X$ of optically active states, i.e.those with  in-plane momentum close to 0 \cite{Combescot_book}. We then performed time and spatially resolved interferometry  in order to quantify $\tau_X$, using a Mach-Zehnder interferometer where the photoluminescence field $\psi$
is recombined with itself after a controlled delay $\tau$ is introduced. Figure 4.b shows the pattern measured above threshold ($D\sim11$) for $\tau$= 5 ps. 
Interference fringes are clearly observed all across the trap and we
 evaluate the interference contrast $|g^{(1)}|(\tau)\sim|\langle \psi^*(t)\psi(t+\tau)\rangle_t|$, $\langle ... \rangle_t$  denoting the average over $t$, by quantifying the modulation amplitude along the vertical axis across the trap (Figure 4.c).

Figure 4.a shows that we observe a clear change of $|g^{(1)}|$($\tau$) at varying phase space densities: Below threshold ($D=3$ in black), the contrast decays exponentially with a characteristic time $\tau_X\sim$4 ps. This decay reflects the two-body collision rate, which is comparable to the value expected in the Thomas-Fermi regime $\Gamma=\hbar\tilde{g}^2 n/2m$. As a result the interference pattern is not resolved for $\tau$ beyond 1/$\Gamma$. By contrast, around threshold ($D\sim7$ in blue) and more markedly above threshold ($D\sim$11 in red)  temporal coherence at long delays $\tau$ is maintained. Remarkably, for $D\gtrsim D_C$ we note that quasi-long range order compatible with an algebraic decay $\sim \tau^{-\eta}$ emerges, as clearly seen Fig.4.a for data plotted in log-log scale. 
As common for systems of mesoscopic sizes, quantitative determination of the algebraic decay rate
will suffer from considerable systematic bias due to the finite range limitations. 
Within our precision, the exponent $\eta$ at criticality is compatible with $\eta_c \simeq 0.25$ predicted for a BKT transition in the thermodynamic limit \cite{KT}.
Averaging the  contrast over a finite region with small density variations may result
 in significantly higher values for the apparent exponent than in strictly homogeneous systems
\cite{Murphy2015,Boettcher2016}. Here, the interference contrast is extracted along one direction in the central density region, making our measurements less sensitive to small density variations,
and thus closer to the homogeneous limit. Finally, we note that emergence of algebraic order has so far been reported only for ultracold atomic gases \cite{Dalibard_2006,Murphy2015} and driven-dissipative exciton-polaritons \cite{Caputo_2018}. Unlike our dipolar fluid, both are distinguished by much weaker interactions so that
the interaction-driven BKT transition competes with Bose-Einstein condensation in trapped
geometries \cite{HCK,Fletcher15}.

To conclude, we have reported evidence for a BKT-type transition in exciton fluids 
with strong dipolar interaction, by observing a pronounced change from exponential to algebraic
decay in the temporal coherence inside a fluctuation region characterized by a universal behaviour
of the excitonic equation of state.
While  our measurements of density fluctuations
 have direct implications for the propagation of first sound, we believe that 
probing second sound by an additional local heating source, e.g. using an auxiliary laser excitation, 
is at experimental reach. In the quest for direct probes of superfluid signatures, second sound measurements open the possibility to observe the discontinuous jump at criticality \cite{Stringari_18}, predicted by the BKT theory, which has only been unambigously observed in Helium films \cite{Reppy_1979}.

\section*{Acknowledgments}
We are grateful to F. Chiaruttini for important discussions regarding the evaluation of the exciton density, and to Alice Sinatra and Yvan Castin for their crucial help concerning the interpretation of the temporal coherence measurements and helpful comments. We also thank C. Lagoin for a critical reading. Our work has been financially supported by the Labex Matisse, the Fondation NanoSciences (Grenoble), and by OBELIX from the French Agency for Research (ANR-15-CE30-0020). The work at Princeton University was funded by the Gordon and Betty Moore Foundation through the EPiQS initiative Grant GBMF4420, and by the National Science Foundation MRSEC Grant DMR 1420541.

\section*{Methods}

$\bullet$ \textit{Sample structure and experimental procedure}

At the core of our sample structure lie two coupled GaAs quantum wells, each being 8 nm wide, separated by a 4 nm Al$_{.3}$Ga$_{.7}$As barrier. The bottom quantum well is positioned 900 nm below the surface of the sample and 150 nm above our sample substrate made of Si-doped GaAs. Electric fields are applied perpendicular to the double quantum well by applying DC potentials between semi-transparent gate electrodes deposited on the surface, and the sample substrate which serves as ground potential. The surface electrodes are patterned in order to realise a 20 $\mu$m electrostatic trap. For that purpose, we designed a  20 $\mu$m diameter disk shaped electrode surrounded by an outer guard gate with 200 nm separation. By applying a larger potential (-4.8V)  onto the disk shaped electrode, compared to its outer guard electrode (-4.3 V),  we realise a 20 $\mu$m  electrostatic trap for dipolar excitons. These are indeed high-field seekers and then remain trapped in the plane of the double quantum well under the central disk electrode.    

As in previous works \cite{Dang_2019,Anankine_2017}, we used a 100 ns long laser excitation in order to inject optically electrons and holes in the two quantum wells, dipolar excitons being formed once carriers have tunnelled towards their minimum energy states, which each lie in a different quantum well. The excitons electric dipole moment then amounts to about 12 C.nm. The laser is set resonant with the direct excitonic absorption of the quantum wells, such that the density of optically injected excess free carriers is minimized. The excitation laser is about 5 $\mu$m  wide, positioned at the center of the electrostatic trap and we typically impose a dead time of at least 100 ns before analyzing the photoluminesence radiated from the trap. Thus, we ensure that our measurements are restricted to the regime where the photo-current is almost entirely evacuated. 

$\bullet$ \textit{Evaluation of the exciton density}

In our studies, the exciton density is extracted from the blueshift of the photoluminescence energy induced by repulsive dipolar interactions between excitons. In practice, we compare the spatial profile of the photoluminescence spectrum at a given density to the one in the very dilute regime to deduce this blueshift. This is for instance directly done by subtracting the profile displayed in Fig. 1.c and 1.b. 

A first crude approach to translate the blueshift of the photoluminescence $\delta E$ into the exciton density $n_X$ is usually referred to as the plate capacitor formulae, which computes the screening of the internal electric field induced by the electron and hole planes confined in a bilayer heterostructure. Assuming a homogeneous density distribution of electronic carriers, one easily deduces with this assumption that $\delta E$=($n_X e^2 d$)/$\epsilon$, where $\epsilon$ is the dielectric constant, $d$ the excitons dipole moment and $e$ the electron charge. However, it has been shown that the plate capacitor formulae largely underestimates the excitons density because it discards the impact of dipolar interactions at short distances \cite{Ivanov_2010,Schindler_2008,Rapaport_2009}. Indeed, these induce a depletion around excitons arising from the energetically prohibitive cost of the dipolar potential at this length scale  

Laikhtman and Rapaport  \cite{Rapaport_2009} have shown that the mean-field treatment yielding the plate capacitor expression can be refined by accounting for the non-uniform distribution induced by the dipolar interaction. Precisely, the authors assume that the probability of finding an exciton at a distance \textbf{r} from another exciton is given by $n_X e^{-U(\textbf{r})/T}d^2\textbf{r}$ where $U(\textbf{r})$ is the repulsive dipolar potential between 2 excitons. Thus, the blueshift of the photoluminescence energy becomes $\delta E$=$\frac{n_X e^2 d}{\epsilon}$$\cdot$$f$ where the screening amplitude $f$ reads
\begin{equation}
f=\Gamma(4/3) \sqrt[3]{\frac{\epsilon\pi d T}{2e^2}} 
\end{equation}

Ivanov et al. \cite{Ivanov_2010} have introduced an alternative approach to evaluate the impact of the screened dipolar repulsions onto the photoluminescence blueshift. The authors have considered a thermally screened  dipolar potential which cuts the bare mid-range $1/r^3$ interaction between excitons. Thus, they evaluate the local density of excitons in quasi-equilibrium (Eq.(1) in \cite{Ivanov_2010}) and compute the factor $f$ correcting the plate capacitor formulae.
\vspace{.5cm}

\centerline{\includegraphics[width=.45\textwidth]{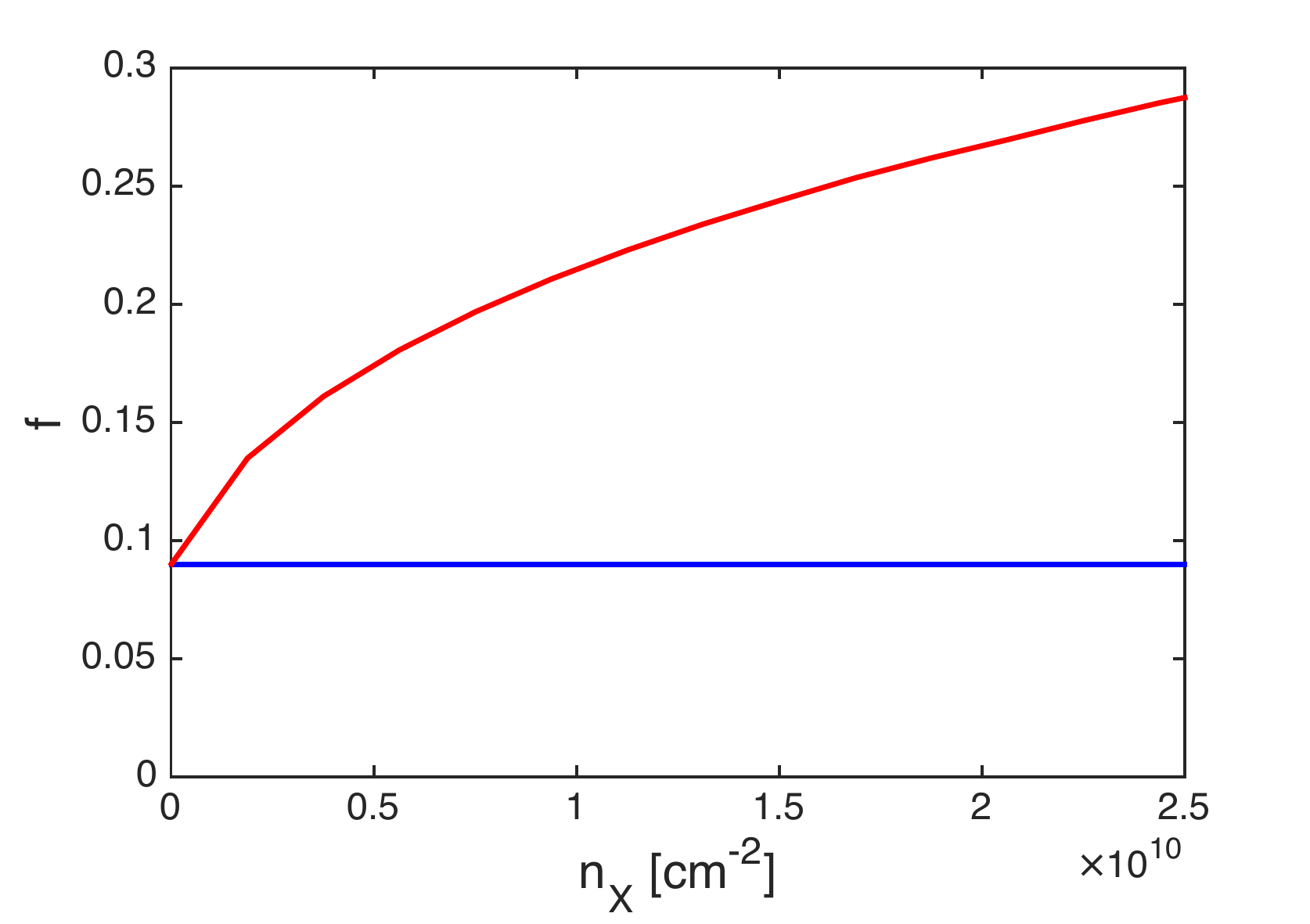}}
\textit{\textbf{Fig. S1}: Screening factor $f$ computed using the model discussed in Refs. \cite{Rapaport_2009} (blue) and \cite{Ivanov_2010} (red).}
\vspace{.5cm}

Fig. S1 compares the amplitude of $f$ for $T$= 1K deduced from Refs. \cite{Ivanov_2010} and \cite{Rapaport_2009}. We then note that the 2 models agree in the very dilute regime, as expected, while Ivanov et al. predict that $f$ is up to 3-fold larger than the prediction of Laikhtman and Rapaport at the largest density we explore ($n_X\sim$ 2 10$^{10}$ cm$^{-2}$). To account for this discrepancy, in Figure 2 we have displayed the variation expected for $\tilde{g}$ obtained by considering both theoretical models.

$\bullet$ \textit{Density flucutations and sound velocity}

Analysing the shot-to-shot density fluctuations gives access to the density-density
pair correlation function, $g(r)$, given by
\begin{equation}
g(r) = \langle n(r) n(0) \rangle
\end{equation}
which is related to the expectation value of the density squared, $\langle n^2 \rangle$, when
integrated over a large volume $V$,
\begin{equation}
\langle n^2 \rangle = \frac{1}{V} \int d{\bf r} g(r)  
\end{equation}
The characteristic distance for the pair particle correlation function is the 
healing length $\xi \sim (2m g n)^{-1/2}/\hbar$, and we can expect
\begin{equation}
\langle n^2 \rangle \approx \frac{1}{\xi^2}\int_0^{\xi^2} d (r^2) g(r)
\end{equation}
Since our spatial resolution exceeds the healing length, the experimental analysis of the shot noise therefore provides a measure of the local density fluctuations, $\sigma^2=\langle n^2 \rangle - \langle n\rangle^2$.

Density fluctuations can be thermodynamically related to the isothermal compressibility, $\kappa_T=-V^{-1} \partial V/\partial P$, via
\begin{equation}
\langle N^2 \rangle - \langle N \rangle^2 =  \frac{\partial^2 \log Z}{\partial (\beta \mu)^2 }
=\frac{\partial \langle N\rangle}{\partial \beta \mu}
=k_B T V n^2 \kappa_T
\end{equation}
or
\begin{equation}
\kappa_T= \frac{V}{k_B T} \frac{\langle n^2 \rangle - \langle n\rangle^2 }{\langle n \rangle^2}
\end{equation}

In the quasi condensate regime which is well described by the Thomas-Fermi approximation of the
density of state, we have $\mu=g n$ or $n^2 \kappa_T=g^{-1}$, so that we obtain 
$\sigma^2 \lambda_T^2 \propto n^2 \kappa_T \propto 1/\tilde{g}$ at constant $V$.
In terms of the compressibility, the speed of sound, $c$, is given by \cite{LandauStat2}
\begin{equation}
\frac1{2m c^2} = \frac12 n \kappa_T
\end{equation}
or $c^2=(m n \kappa_T)^{-1} = g n/m $.

The rate of inelastic two-body collisions is given by $2 \Gamma=m g^2 n/\hbar^3$ \cite{Petrov} which in the
Thomas-Fermi limit gives
\begin{equation}
\Gamma=\frac{\hbar \tilde{g}^2n}{2m} \approx \frac{m \tilde{g} c^2}{\hbar} 
\end{equation}
The time scales indicate that the experimental measured broadening $\Gamma_{exp}/n$ before
the onset of algebraic order is of order of
the inelastic two-body collisions.


\begin{thebibliography}{99}

 \bibitem{Berenzinskii} V. L. Berezinski{\v i}, Sov. Phys. JETP {\bf 34}, 610 (1972).

\bibitem{KT} J. M. Kosterlitz and D. J. Thouless, J. Phys. C: Solid State Phys.
6, {\bf 1181} (1973); J. M. Kosterlitz, J. Phys. C: Solid State Phys. {\bf 7}, 1046
(1974).

\bibitem{Mermin-Wagner} N.D. Mermin and H. Wagner, Phys. Rev. Lett. {\bf 17}, 1133 (1966).

\bibitem{Combescot_ROPP} M. Combescot, R. Combescot, F. Dubin, Rep. Prog. Phys. \textbf{80}, 066401 (2017).
 
 \bibitem{Maciej_2011} C Trefzger, C Menotti, B Capogrosso-Sansone, and M Lewenstein, J. of Phys. B: Atomic, Mol. and Opt. Phys. \textbf{44}, 17 (2011)

 \bibitem{Filinov10} A. Filinov, N. V. Prokof'ev, and M. Bonitz, Phys. Rev. Lett. {\bf 105}, 070401 (2010).

 
\bibitem{Tanzi_19} L. Tanzi et al., Phys. Rev. Lett. \textbf{122}, 130405 (2019)

\bibitem{Bottcher_19} F. Bottcher et al., Phys. Rev. X \textbf{9}, 011051 (2019)

\bibitem{Chomaz_19} L. Chomaz et al., Phys. Rev. X \textbf{9}, 021012 (2019)

\bibitem{Combescot_book} M. Combescot and S.Y. Shiau "Excitons and Cooper Pairs: Two Composite Bosons in Many-Body Physics" (Oxford. Univ. Press, 2016)


\bibitem{High_2012} A. High et al., Nature \textbf{483}, 584 (2012)


\bibitem{Anankine_2017} R. Anankine et al., Phys. Rev. Lett. \textbf{118}, 127402 (2017)

\bibitem{Nandi_2012} D. Nandi et al., Nature \textbf{488}, 481 (2012)

\bibitem{Dang_2019} S. Dang et al, Phys. Rev. Lett. \textbf{122}, 117402 (2019)

\bibitem{Rapaport_2019} Y. Mazuz-Harpaz et al., PNAS \textbf{10}, 116 (2019)


\bibitem{Yefsah_2011} T. Yefsah, R. Desbuquois, L. Chomaz, K.J. Guenter, J. Dalibard, Phys. Rev. Lett. \textbf{107}, 130401 (2011); 
R. Desbuquois, T. Yefsah, L. Chomaz, C. Weitenberg, L. Corman, S. Nascimbène, and J. Dalibard, Phys. Rev. Lett. \textbf{113}, 020404 (2014).

\bibitem{SaintJalm19} R. Saint-Jalm, P. C. M. Castilho, {\'E}. Le Cerf, B. Bakkali-Hassani, J.-L. Ville, S. Nascimbene, J. Beugnon, and J. Dalibard, Phys. Rev. X \textbf{9}, 021035 (2019).

\bibitem{Hung_2011} C.L. Hung, X. Zhang, N. Gemelke, C. Chin, Nature \textbf{470}, 236 (2011)

\bibitem{Prokofev2002} N. Prokof'ev and B. Svistunov, Phys. Rev. A {\bf 66}, 043608 (2002).

\bibitem{Nelson} D.R. Nelson and J.M. Kosterlitz, Phys. Rev. Lett. {\bf 39}, 1201 (1977).


\bibitem{Ivanov_2010} A. L. Ivanov, E. A. Muljarov, L. Mouchliadis, and R. Zimmermann, Phys. Rev. Lett. \textbf{104}, 179701 (2010).

\bibitem{Schindler_2008} C. Schindler and R. Zimmermann, Phys. Rev. B \textbf{78}, 045313 (2008).

\bibitem{Rapaport_2009} B. Laikhtman and R. Rapaport Phys. Rev. B \textbf{80}, 195313 (2009).

\bibitem{Popov} V. N. Popov, Functional Integrals in Quantum Field Theory
and Statistical Physics (D. Reidel Publishing, Dordrecht, 1983).

\bibitem{HCK} M. Holzmann, M. Chevallier, and W. Krauth,
Phys. Rev. A {\bf 81}, 043622 (2010).

\bibitem{Pnas} M. Holzmann, G. Baym, J.-P. Blaizot, and F. Laloe,
Proc. Natl. Acad. Sci. USA, 10.1073/pnas.0609957104 (2007).

\bibitem{Fletcher15} R. J. Fletcher, M. Robert-de-Saint-Vincent, J. Man, N. Navon, R. P. Smith, 
K. G. H. Viebahn, and Z. Hadzibabic, Phys. Rev. Lett. \textbf{114}, 255302 (2015). 

\bibitem{Amo_2009} A. Amo et al., Nat. Phys. \textbf{5}, 805 (2009)

\bibitem{Prokofev2018} N.V. Prokof'ev and B.V. Svistunov,  JETP {\bf 127}, 860 (2018).

\bibitem{Caputo_2018} D. Caputo et al., Nat. Mat. \textbf{17}, 145 (2018).


\bibitem{Boettcher2016} I. Boettcher and M. Holzmann, Phys. Rev. A {\bf 94}, 011602(R) (2016).

\bibitem{Murphy2015} P.A. Murthy, I. Boettcher, L. Bayha, M. Holzmann, D. Kedar, M. Neidig, M.G. Ries, A.N. Wenz, G. Zuern, and S. Jochim, Phys. Rev. Lett. {\bf 115}, 010401 (2015).

\bibitem{Dalibard_2006}  Z. Hadzibabic, P. Krueger, M. Cheneau, B. Battelier,  J. Dalibard, Nature \textbf{441}, 1118 (2006)


\bibitem{Stringari_18} M. Ota and S. Stringari, Phys. Rev. A \textbf{97}, 033604 (2018)


\bibitem{Reppy_1979} D.J. Bishop and J. D. Reppy, Phys. Rev. Lett. \textbf{40}, 1727 (1978)

\bibitem{LandauStat2} E. M. Lifshitz and L. P. Pitaevskii, Statistical Physics
(Pergamon Press, Oxford, 1980), Pt. 2.

\bibitem{Petrov} D. S. Petrov and G. V. Shlyapnikov, Phys. Rev. A {\bf 64}, 012706 (2001).

\end{thebibliography}
\end{document}